\documentclass[prl,twocolumn,twoside,showpacs,byrevtex,superscriptaddress]{revtex4}

\usepackage{graphicx,epsfig,verbatim,enumerate}
\usepackage{amssymb,amsmath}
\usepackage{ifthen}
\newboolean{twocolswitch}

\newcommand{\sindex}[1]{}
\newcommand{\nindex}[1]{}

\newcommand{\www}[1]{\url{#1}}

\usepackage{color}

\newcommand{\phiiactive}{\phi_{i,t}}

\newcommand{\phiiup}{{\phi_{i,\rm on}}}
\newcommand{\phiidown}{{\phi_{i,\rm off}}}

\newcommand{\phiup}{{\phi_{\rm on}}}
\newcommand{\phidown}{{\phi_{\rm off}}}

\newcommand{\nodeinfprob}{\phi}

\newcommand{\avgdegree}{k_{\rm avg}}

\newcommand{\stateA}{S_{0}}
\newcommand{\stateB}{S_{1}}

\setboolean{twocolswitch}{true}

\begin{document}

\title{
  Limited Imitation Contagion on Random Networks: Chaos, Universality, and Unpredictability

}

\author{
\firstname{Peter Sheridan}
\surname{Dodds}
}

\email{peter.dodds@uvm.edu}

\affiliation{Department of Mathematics \& Statistics,
  Computational Story Lab,
  Vermont Complex Systems Center,
  \& the Vermont Advanced Computing Core,
  The University of Vermont,
  Burlington, VT 05401.}

\author{
\firstname{Kameron Decker}
\surname{Harris}
}

\email{kameron.harris@uvm.edu}

\affiliation{Department of Mathematics \& Statistics,
  Computational Story Lab,
  Vermont Complex Systems Center,
  \& the Vermont Advanced Computing Core,
  The University of Vermont,
  Burlington, VT 05401.}

\author{
\firstname{Christopher M.}
\surname{Danforth}
}
\email{chris.danforth@uvm.edu}

\affiliation{Department of Mathematics \& Statistics,
  Computational Story Lab,
  Vermont Complex Systems Center,
  \& the Vermont Advanced Computing Core,
  The University of Vermont,
  Burlington, VT 05401.}

\date{\today}

\begin{abstract}
  We study a family of binary state, socially-inspired contagion models
which incorporate imitation limited by an aversion
to complete conformity.  
We uncover rich behavior in our models whether operating 
with either probabilistic or deterministic
individual response functions
on both dynamic and fixed random networks.
In particular, we find significant variation in
the limiting behavior of a population's 
infected fraction, ranging from steady-state to chaotic.
We show that period doubling arises
as we increase the average node degree, 
and that the universality class of this
well known route to chaos depends on the 
interaction structure of random networks 
rather than the microscopic behavior of individual nodes.
We find that increasing the fixedness of the system tends to stabilize
the infected fraction, yet disjoint, multiple equilibria are possible depending
solely on the choice of the initially infected node.
  
\end{abstract}


\maketitle

The structure and dynamics of real, complex networks remains
an open area of great research interest, particularly
in the realm of evolutionary processes acting on and within
networked systems~\cite{newman2003a,watts2007a,colizza2007a,wuchty2007a,hidalgo2007a,liu2011a}.
Here, motivated by considerations of 
social contagion---the spreading of ideas and behaviors
between people through social networks and media---we explore
an idealized, binary-state social contagion model
in which individuals choose to be like others but only
up to a point: they do not want to be 
like everyone else~\cite{simmel1957a,granovetter1986a,ugander2012a,coupledmaps}.  
We term such behavior `Limited Imitation Contagion.'
We build naturally on previous studies
of threshold models of contagion~\cite{schelling1971a,granovetter1978a,watts2002a,gleeson2008a},
and our model can also be seen as a specific
subfamily of dynamical Boolean network models~\cite{kauffman1969a,aldana2003a}.
We show how macroscopic network structure overrides
microscopic details, and we 
find complex dynamics whose character moves 
from universal and predictable to 
particular and unpredictable as 
we allow the system to become increasingly deterministic.

In constructing our model, 
our main interest is in understanding how spreading by Limited Imitation
Contagion on random networks behaves under three main tunable conditions: 
(1) Social awareness: the rate of contact between individuals;
(2) Social variability: the extent to which friendships are fixed;
and
(3) Social influence: the character of individuals' responses to the behavior of others.

To begin with, we consider a binary state model
for which individuals are either in a base state $\stateA$ or an alternate state $\stateB$.
We assume individuals interact over an uncorrelated random network,
which may be dynamic or fixed.
For simplicity, and due to the richness of the dynamics we find,
we employ standard Erd\"{o}s-R\'{e}nyi networks
which possess Poisson degree distributions.
We take time to be discrete ($t=0,1,2,\ldots$),
and we prescribe each node's degree $k$ at $t$$=$$0$.
In a dynamic network, when node $i$ updates, it samples 
the states of $k_i$ randomly chosen nodes
(i.e., the system is a random mixing model with 
non-uniform contact rates).
For a fixed network, node $i$ repeatedly samples the 
same $k_i$ nodes.
We further restrict our attention to single-seed contagion processes
wherein all nodes are in state $\stateA$ at time $t$$=$$0$,
with one randomly chosen node in state $\stateB$.

The contagion process is manifested through the response functions of
individual nodes.  
We allow nodes to update synchronously, and
node $i$'s response function $F_i: [0, 1] \mapsto [0, 1]$ 
gives the probability that node $i$ will be in state $\stateB$ upon
updating, where the argument taken by $F_i$ is the fraction of nodes
sampled by node $i$ that are currently in state $\stateB$, $\phiiactive$.

\begin{figure}[tbp]
  \centering
  \includegraphics[width=\columnwidth]{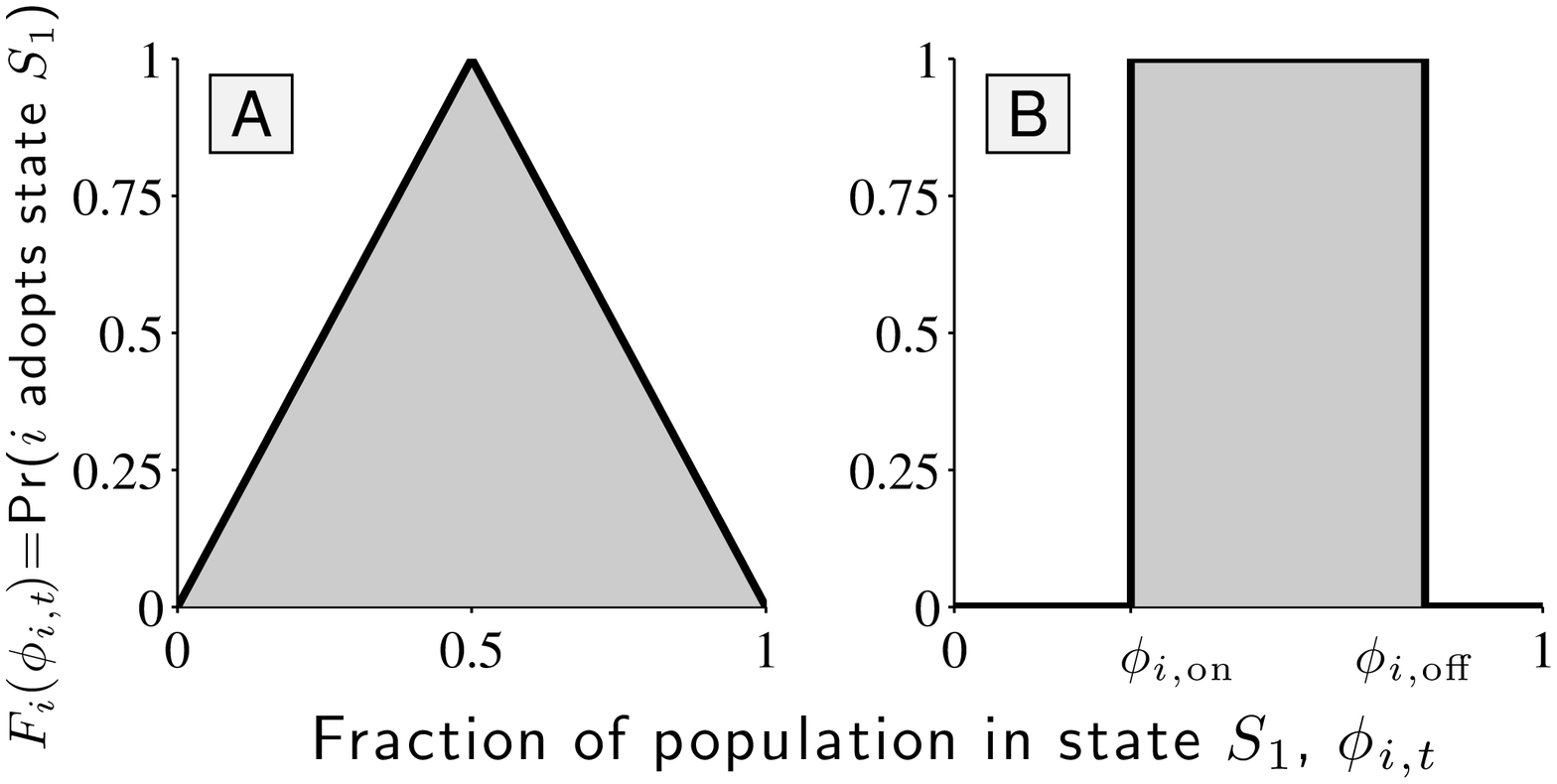} 
  \caption{
    Examples of probabilistic and deterministic response
    functions capturing Limited Imitation Contagion dynamics.
    At each time $t$, nodes use their given response functions
    to update their own state based on the perceived fraction
    of their neighbors in state $\stateB$, $\phiiactive$.
    We construct the tent map $T_2$ (see main text) shown
    in (A) by averaging over deterministic response functions
    of the kind shown in (B)
    by considering a family of the latter with `on' and `off' thresholds
    uniformly distributed in $[0,\frac{1}{2}]$ and $[\frac{1}{2},1]$
    respectively.
    We then build networked systems whose macroscopic character
    is tent map-like but differ strongly at the microscopic level.
    }
  \label{fig:updownrfn_network02}
\end{figure}

We investigate two kinds of response functions,
probabilistic and deterministic, both of which incorporate
the characteristic of the adoption probability growing
and then diminishing as the perceived popularity of $\stateB$ increases.
In Fig.~\ref{fig:updownrfn_network02}A,
we show an example of a probabilistic response function,
the tent map, which is defined as
$
T_r(x) = 
rx 
$
for
$
0 \le x \le \frac{1}{2}
$
and
$
r(1-x)
$
for
$
\frac{1}{2} \le x \le 1.
$
Here, we consider the $r=2$ case
meaning node $i$ adopts state $\stateB$
with probability $T_2(\phiiactive)$.
We use the tent map $T_2$ for a number of reasons:
(1) As a standard iterative map of the unit interval, the tent map's dynamics
for $r=2$ are both interesting 
and well understood~\cite{alligood1996a}---it 
is fully chaotic and its invariant density $\chi$ is uniform on $[0,1]$
($\chi$ is the long term probability distribution for the values of a map's iterates);
(2) The tent map captures a probabilistic flavor of the 
adopt-when-novel, drop-when-ubiquitous behavior we aim to model;
and
(3) We can construct a simple and elegant connection with
deterministic contagion processes which we describe next.

In Fig.~\ref{fig:updownrfn_network02}B,
we show an example of a deterministic response function
which is characterized by `on' and `off' thresholds,
$\phiup$ and $\phidown$.  Node $i$ will only adopt or
remain in state $\stateB$ if it perceives the fraction of
others in state $\stateB$ to lie between its on and off thresholds:
$\phiiup \le \phiiactive < \phiidown$.

For these deterministic `on-off' response functions,
we examine the special case where
$\phiup$ is distributed uniformly on
$[0,\frac{1}{2}]$ and $\phidown$ likewise on $[\frac{1}{2},1]$.
Averaging over all deterministic response functions created
in this way, we obtain precisely the tent map $T_2$.
Collectively, we then have the same on-average behavior
in both the probabilistic and deterministic cases, and 
this allows us to profitably explore the effect of varying the 
specific response dynamics at the micro level,
as well as interaction patterns.

\begin{figure*}[htbp]
\centering
\includegraphics[width=\textwidth]{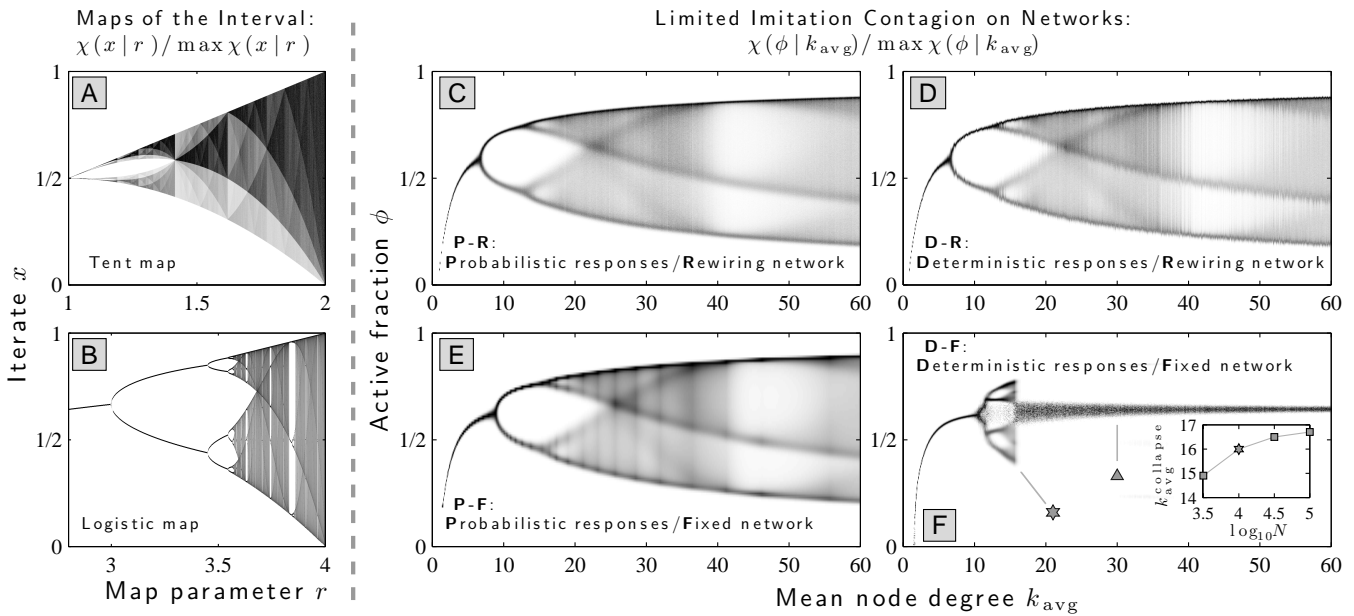}
\caption{
  Comparison of bifurcation diagrams for networked systems with 
  Limited Imitation Contagion
  based on tent-map response functions.
  (A) and (B): For reference, standard bifurcation diagrams for the tent map
  and the logistic map operating as maps of the unit interval,
  showing distinct universality classes.
  The gray scale represents the
  normalized invariant density $\chi$ (darker indicating higher values).
  (C--F): Orbit diagrams for Limited Imitation Contagion acting
  on standard random networks 
  for four model classes with
  probabilistic or deterministic responses and network 
  interactions
  as indicated.
  The tunable parameter is average degree $\avgdegree$.
  The inset in F shows the approximate location of the collapse point 
  $k_{\rm avg}^{\rm collapse}$ as
  a function of network size,
  the stars indicating the collapse point for $N$=$10^4$.
  The triangle in F indicates $\avgdegree=30$ for the system examined
  in Fig.~\ref{fig:chcon.collectedgoodies}.
  Simulation details: 
  network size $N$=$10^4$; 
  one random seed per network;
  granularity of 0.1 in $\avgdegree$;
  binning size of 0.001 in $\nodeinfprob$;
  for P-R, D-R, and P-F systems, we ran simulations for $10^4$ time steps
  on $10^3$ networks,
  taking values of $\nodeinfprob_{t}$ for $t \ge 10^3$;
  and for D-F systems, which exhibit long transient behavior (see text),
  we ran simulations for $10^5$ time steps
  on 200 networks, ignoring the first $5\times 10^4$ time steps.
}
\label{fig:chcon.bifdiags}
\end{figure*}

We now examine the behavior of four subfamilies of our model
that vary in terms of node response functions
being probabilistic or deterministic (P or D),
and whether or not network connections 
randomly rewire or are fixed (R or F).
We refer to these model classes as P-R, P-F, D-R, and D-F.
Each class is indexed by the parameter of average node degree $\avgdegree$,
and, as per our design, all four systems have the same
on-average response functions (the tent map $T_2$) 
and degree distributions (Poisson).

Our first focus is on the long term behavior of these
four networked systems.
Key to our understanding are the well-known behaviors 
of the tent and logistic maps
acting iteratively on the unit interval,
the former given above and
the latter by $L_r(x) = rx(1-x)$.
We reproduce the orbit diagrams for these models
in Figs.~\ref{fig:chcon.bifdiags}A and B.
Both systems are controlled by the amplitude parameter $r$,
whose increase leads to changes in their invariant densities,
famously resulting in the bifurcation diagram for the logistic map.
The two maps, while topologically conjugate, 
produce distinct bifurcation diagrams.

As we show in Fig.~\ref{fig:chcon.bifdiags}C--F,
the signature of 
the microscopic response function of the tent map is erased
by the network dynamics of the four model classes.
Macroscopically, we see four orbit diagrams 
analogous to the logistic map's
characteristic bifurcation diagram.
We see that increasing the average connectivity
of the network $\avgdegree$ is equivalent to 
increasing the logistic map's amplitude parameter $r$,
and the system moves along the period-doubling
route to chaos.  

However, while appearing to belong to the same universality class,
the orbit diagrams of the four models differ 
importantly in detail, most profoundly 
for the fully deterministic D-F class.

First, we observe that the four systems produce
only three distinct orbit diagrams,
since for large enough systems, 
the two random mixing classes P-R and D-R
must exhibit the same macroscopic behavior (Figs.~\ref{fig:chcon.bifdiags}C and D).
For the D-R class, random rewiring overwhelms
the fact of each node having a fixed deterministic response function
(for finite systems some evidence of discreteness limits
the smoothness of the orbit diagram in Fig.~\ref{fig:chcon.bifdiags}D).
While we present simulation results only here,
we note that we are able to fully address the P-R and D-R models analytically,
and we find excellent agreement (see~\cite{harris2012b} where we explore more general influence maps).

In next considering the fixed network model
with probablistic response functions, P-R,
we see in Fig.~\ref{fig:chcon.bifdiags}E
that the orbit diagram has 
slightly moved to the right---the bifurcation points
now occur for slightly higher values of $\avgdegree$
relative to the random mixing cases.
A robust, discernible vertical striation also appears, with separation
between stripes increasing with $\avgdegree$, for which we do not
have an explanation.
Thus, making the network fixed appears to induce
some modest changes in the dynamics, though we 
cannot discount finite size effects.

Our final model class, D-F---to whose description we devote the remainder of the paper---is the most structured, 
and it generates considerably different and intricate behavior.
Once the network and set of response functions is realized, 
the system is now completely deterministic (and irreversible),
and we find surprising changes in the system's behavior at both macroscopic and microscopic scales.
In Fig.~\ref{fig:chcon.bifdiags}F, we see 
that the orbit diagram for the  D-F model 
collapses abruptly after several rounds of period doubling,
which are themselves relatively compressed in terms of $\avgdegree$.
Our simulations
suggest that above an average degree of $\avgdegree\!\simeq\!17$,
the macroscopic dynamics always collapse.  
The inset in Fig.~\ref{fig:chcon.bifdiags}F
shows estimates of the collapse point for $N$$=$$10^{3.5}$ up to $N$$=$$10^5$,
and the asymptotic growth with $\log_{10}N$ strongly suggesting the collapse is real
and not a finite size effect.
The collapse appears to favor a fixed point macroscopic state, around 
$\nodeinfprob$$=$$2/3$, which is the fixed point of the tent map $T_2$.
However, a closer examination of the D-F class's potential dynamics
reveals a far more subtle story.  

\begin{figure*}[htbp]
  \centering
  \includegraphics[width=\textwidth]{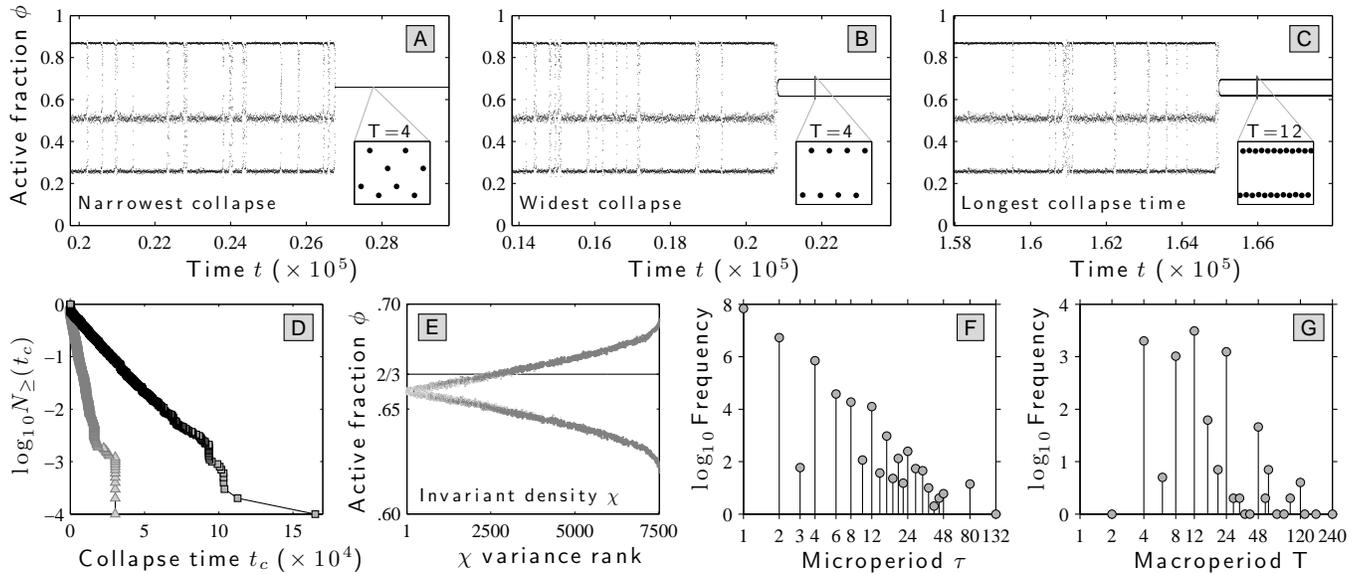}
  \caption{
    Comprehensive survey of the possible 
    Limited Imitation Contagion dynamics for a single fixed random network
    ($\avgdegree$=$30$) with its $N$=$10^4$ nodes having fixed deterministic response functions.
    We ran $N$ simulations, activating each node as a single seed
    and allowed the fully deterministic
    dynamics to run until collapse.
    (A-C): Partial time series for three different initial 
    seeds leading to the narrowest and widest post-collapse dynamics (A and B)
    and the longest collapse time $t_c$ (C).
    The insets in A, B, and C magnify two cycles of the underlying
    periodic behavior.
    (D): Complementary cumulative distribution of collapse times 
    (black squares)
    compared with an example distribution from another network (gray triangles);
    (E): Invariant densities, ordered by variance;
    (F): Histogram of post-collapse microperiod $\tau$;
    (G): Histogram of post-collapse macroperiod $T$.
      }
  \label{fig:chcon.collectedgoodies}
\end{figure*}

In Fig.~\ref{fig:chcon.collectedgoodies},
we summarize the possible dynamics of a lone realized network 
($N$=$10^4$, $\avgdegree$=$30$)
with a single set of fixed deterministic response functions.
We exhaustively ran $N$$=$$10^4$ tests of the system's behavior by separately seeding
each individual node.  
Of these, 7515 contagion events successfully
lead to long-term, non-zero infection levels.
Figs.~\ref{fig:chcon.collectedgoodies}A--C show partial time series
for three system evolutions for the same network, differing only by seed node,
and corresponding respectively to the narrowest and widest collapses,
as measure by the variance of $\chi$,
and the longest time to collapse.
In each case, the initial dynamics of $\nodeinfprob$
exhibit a clear period three pattern interspersed with intermittent
chaotic dynamics, followed by a sharp collapse to distinct periodic behaviors.

Across all initial seeds,
the time to collapse $t_c$ varies greatly, with the extreme example of 
Fig.~\ref{fig:chcon.collectedgoodies}C
collapsing after more than 160,000 time steps and over 1.6$\times$$10^9$ individual node updates.
In Fig.~\ref{fig:chcon.collectedgoodies}D, we provide the complementary
cumulative distribution for collapse times (black squares).
The semi-log scale indicates that an expontial decay for $t_c$$<$$10^5$
covers the majority of cases with the longest collapse time clearly
a singular outlier.  
Even after the dynamics collapse, we see that on the order of
10\% of all nodes change state in each time step.
Surprisingly, we see different distributions for $t_c$ for
different individual networks.
For example, the grey triangles represent the collapse time distribution
for another randomly realized  network with $\avgdegree$=$30$.
Again we see an exponential distribution but now with a much
steeper decay.  Rather than finding one outlier,
we see that a set of exceptional dynamics
lasts for much longer than the main exponential decay would suggest, 
with an isolated set distributed around $t_c \simeq$ 25,000,
and a group ending at around $t_c \simeq$ 30,000.

We turn lastly to the behavior after the collapse.  
We call the post-collapse period of $\nodeinfprob_t$ the system's `macroperiod' $T$,
and the period of an individual node its `microperiod' $\tau$.
Fig.~\ref{fig:chcon.collectedgoodies}E
shows the invariant density $\chi$ for all seeds
post collapse, ordered by $\chi$'s variance
(contagion events that fail are not included).
Strikingly, different seeds almost always lead to distinct invariant densities,
with an apparent dominance of period $T$=2 behavior (but see below).
We also find that the tent map's fixed point of 
2/3 (horizontal line in Fig.~\ref{fig:chcon.collectedgoodies}E)
is not the center of the invariant densities.

In Figs.~\ref{fig:chcon.collectedgoodies}F and G,
we provide complete histograms of all post-collapse
microperiods and macroperiods for this particular network
(again ignoring failed spreading events).
Unlike the collapse time distribution,
we see broadly similar distributions for other individual networks.
Microperiods such as 1, 2, 4, 12, and 24 form a dominant envelope,
and many odd-numbered microperiods are absent.
We find that the most common resultant macroperiods are 4, 8, 12, and 24;
that a pure macroperiod 2 is relatively rare ($<$ 0.1\%);
and that the largest observed macroperiod is 240.
Again, all of these outcomes are deterministic, depending
only on the choice of initial seed.

To conclude, in abstracting from a real world problem, 
we have constructed a rich, networked-based model 
of Limited Imitation Contagion
that allows us to deeply explore the effects of
manifesting either deterministic or probabilistic microscopic behavior
in both network structure and node response in a well-defined fashion.  
Our findings broadly suggest that increasing the degree of a social system's microscopic rigidity
leads to higher levels of long-term unpredictability at macroscopic scales.
Future work could take at least two paths, one directed toward the dynamical systems aspects
and the other toward a more realistic empirically supported model.
Some natural objectives would be to explain why the collapse occurs for the D-F class;
to explore how local properties of seed nodes such as node degree,
correlations, and clustering relate to the collapse time and post-collapse dynamics;
to test other contagion mechanisms across systems of
increasingly fixed microscopic structure;
and to examine these Limited Imitation Contagion models on 
real networks and classes of naturally occurring ones.

\acknowledgments
The authors thank A. Mandel and D. J. Watts for encouragement, 
and are grateful for the computational resources provided by 
the Vermont Advanced Computing Core supported by NASA (NNX 08A096G).
KDH was supported by VT-NASA EPSCoR;
CMD and PSD were supported by a grant from the MITRE Corporation;
PSD was supported by NSF CAREER Award \#0846668.

\end{document}